# A Metalens with Near-Unity Numerical Aperture

Ramón Paniagua-Domínguez[1†]*, Ye Feng Yu[1†], Egor Khaidarov[1, 2†], Sumin Choi[1], Victor Leong[1], Reuben M. Bakker[1], Xinan Liang[1], Yuan Hsing Fu[1], Vytautas Valuckas[1], Leonid A. Krivitsky[1] and Arseniy I. Kuznetsov[1#]

[1]Data Storage Institute (Agency for Science, Technology and Research, A*STAR), 2 Fusionopolis Way, #08-01, Innovis 138634, Singapore.

[2]LUMINOUS! Center of Excellence for Semiconductor Lighting and Displays, The Photonics Institute, School of Electrical and Electronic Engineering, Nanyang Technological University, 50 Nanyang Avenue, 639798, Singapore

*Correspondence to: ramon-paniag@dsi.a-star.edu.sg.

#Correspondence to: arseniy_k@dsi.a-star.edu.sg.

†These authors contributed equally to this work.

## Abstract

The numerical aperture (NA) of a lens determines its ability to focus light and its resolving capability. Having a large NA is a very desirable quality for applications requiring small light-matter interaction volumes or large angular collections. Traditionally, a large NA lens based on light refraction requires precision bulk optics that ends up being expensive and is thus also a specialty item. In contrast, metasurfaces allow the lens designer to circumvent those issues producing high NA lenses in an ultra-flat fashion. However, so far, these have been limited to numerical apertures on the same order of traditional optical components, with experimentally reported values of NA <0.9. Here we demonstrate, both numerically and experimentally, a new approach that results in a diffraction limited flat lens with a near-unity numerical aperture (NA>0.99) and sub-wavelength thickness (~$\lambda$/3), operating with unpolarized light at 715 nm. To demonstrate its imaging capability, the designed lens is applied in a confocal configuration to map color centers in sub-diffractive diamond nanocrystals. This work, based on diffractive elements able to efficiently bend light at angles as large as 82°, represents a step beyond traditional optical elements and existing flat optics, circumventing the efficiency drop associated to the standard, phase mapping approach.

The pioneering work of Ernst Abbe established a lower limit to the size of a spot, $\Delta$, that can be created by focused light. The spot size is determined by the wavelength of light, $\lambda$, the refractive index of the medium in which it is traveling, $n$, and the angle under which it is focused, $\theta$, through the formula[1]:

$$\Delta = \frac{\lambda}{2n\sin\theta} \qquad (1)$$

The quantity $n\sin\theta$ is often referred to as the numerical aperture, NA. In the case of a lens, equation (1) sets the ultimate resolution that can be achieved when it is free from all manufacturing imperfections; this is, when it is only limited by diffraction. In that case, $\theta$ represents the maximum acceptance angle of the lens. It is related to its focal distance, $f$, and the size of the lens, $d$, by the equation[1] $\operatorname{tg}\theta = d/2f$. An alternative definition of the numerical aperture, derived from the point spread function, stablishes that the NA is connected to the Airy radius $r_A$, defined as the radius from the central peak of the Airy function to its first minimum, via the relation[1] $r_A = 1.22\lambda/2\text{NA}$. Using reciprocity arguments, the NA of the lens also determines its capability to collect light emitted from a point source placed at its focus. Therefore, a large NA is also very desirable in situations when low-light signals are involved, and collecting as many photons as possible is extremely valuable. Examples of such situations are quantum optics systems involving the emission of single photons with random angular distributions or microscopy involving low photon numbers. Unfortunately, traditional large NA objectives are quite large and complex, which limits their practical use and may constrain the scalability of the system. In addition, currently, the maximum free-space NA of optical objectives available in the market is typically limited to 0.9-0.95, corresponding to a collection angle $\theta < 72°$.

For the last two decades, different approaches have been explored to substitute traditional optical components by flat optical ones. Two particularly successful ones have been the so-called "échellete" blazed gratings[2], with some limitations due to the shadowing effect, and the blazed binary gratings[3-5], which overcame those. The interested reader can refer to a recent review to see a "historical fresco" on the development of these fields[6]. In both cases, the height of the patterned structures is typically larger or, in the best cases, equal to the wavelength of the incoming light, which makes fabrication of large area devices based on such components particularly challenging. If the sub-wavelength thickness is desired, one possibility is to use resonant structures, which naturally provide additional phase accumulation that can be used to achieve the full $2\pi$ phase mapping[7, 8]. These concepts were first demonstrated using resonant plasmonic elements for a wide range of applications, including lenses[9, 10], vortex beam generation[11] or holography[12]. However, their applicability to realize efficient transmission devices is limited, due to the associated Ohmic losses and strong reflections. An alternative approach to realize these devices uses high refractive index dielectrics to replace metallic nanostructures[13-16]. As an example, high-contrast transmit-arrays employing non-resonant nanostructures with sizes comparable to the wavelength have demonstrated remarkable efficiencies (>80%) for light focusing and holography[17-20]. Apart from their obvious advantages in terms of lower dissipative losses, the possibility to excite more than one resonance simultaneously in dielectric particles[21-24] opened new pathways in the field, such as the realization of Huygens' metasurfaces with full transmission and $2\pi$ phase control with truly sub-wavelength ($< \lambda/5$) structures[25-28]. Using some of these concepts, several works have recently demonstrated flat lenses with focusing characteristics comparable to commercial ones[19, 29-30]. So far, however, none of them have demonstrated numerical apertures above those of commercially available microscope objectives, being the largest ever reported ~0.87 for a free space configuration [19] (~1.1 for an oil-immersed one[31]), which corresponds to a maximum collection angle of around 60°. In this work, we report a flat lens design with truly sub-wavelength thickness ($\sim\lambda/3$) and near-unity NA that far exceeds any previously reported values for flat and bulk optics, namely an NA>0.99 (corresponding to a collection angle of 82°) at 715 nm. This lens, which significantly expands the range of collection angles as compared to any other previously reported, is then used in a confocal configuration to map the emission from color centers in a sub-diffractive diamond nanocrystal and demonstrate its imaging capabilities.

## Results

Our lens works on the basis of diffracted energy redistribution by means of nanoantenna inclusions with asymmetric scattering patterns. This novel concept does not rely on the traditional phase mapping approach and, thus, is capable of circumventing its inherent drop in efficiency when a small number of elements is used to map the phase profile. As we show below, this new perspective to tackle the problem allows efficient bending of transmitted light at angles up to 82°. Recently, this problem of bending electromagnetic waves at large angles using metasurfaces has attracted significant attention, with several works reporting remarkably high efficiencies by simultaneous continuous modulation of both the local phase and the impedance of the wave[32-35]. This approach is, however, challenging to implement in the optical spectrum, for which the constituent elements are not deeply sub-wavelength, thus facing the same discretization issues as mentioned before. On the other hand, alternative attempts to optimize the efficiency using genetic algorithms have proven successful[36-38]. However, limited physical insight is obtained using this method, and no general design rules can be, thus, extracted. Here, instead, we propose an alternative approach relying on the intuitive concepts of phased array antennas and tailored scattering patterns, which allows a judicious design of the systems and that can be experimentally realized with conventional nanofabrication approaches[39].

*Large-angle light bending with arrays of nanoantennas with tailored scattering patterns.*

The operating principle, schematically depicted in Fig. 1a, is as follows. Consider a simple square lattice diffracting an incident light beam. By adjusting the periods of the unit cell one can select the number of diffraction orders and their corresponding angles. In this particular case, let us choose one of the periods, $P_d$, to be diffractive for transmitting and reflecting light, up to the first order, under normal incidence illumination and generate a set of diffraction orders such that $T_{+1}$ corresponds to the desired bending direction. The other period, $P_{nd}$, is chosen to be non-diffractive. We now consider that each unit cell of the lattice contains a nanoantenna whose scattering pattern can be designed to selectively redistribute the transmitted energy among the diffraction orders. In particular, we attempt to concentrate as much energy as possible into the desired bending direction ($T_{+1}$ order). To do so, the scattering from each nanoantenna must be inhibited in the directions of the other diffraction orders that are opened, i.e. $T_{-1}$, $R_0$, $R_{+1}$ and $R_{-1}$, as schematically shown in Fig. 1a. In the direction of the zeroth-order in transmission, $T_0$, the scattering must compensate the field transmitted in the absence of the nanoantennas, that is, to have the same amplitude and opposite phase. Note that scattering at all other angles except of those corresponding to the diffraction order directions is compensated by the rest of the unit cells in the array. If the above conditions are simultaneously met, and in the absence of absorption, all incident light will be channeled into the desired $T_{+1}$ diffraction order with 100% efficiency, causing the light to bend at the angle corresponding to the first order of diffraction, which can be controlled by the array period.

Figure 1b shows the unit cell of an array designed to bend normally incident light (715 nm wavelength) at an angle of 82°. It consists of an asymmetric dimer nanoantenna made of silicon (Si) nanodisks of height $H = 250$ nm placed on top of a silicon dioxide ($SiO_2$) substrate and surrounded by air. The cylinder diameters are $D_1 = 150$ nm and $D_2 = 190$ nm and the gap $g = 50$ nm. The diffractive and non-diffractive periods of the square lattice are, respectively, $P_d = 721$ nm and $P_{nd} = 260$ nm. The asymmetric dimer nanoantennas have been previously shown to generate asymmetric radiation patterns even in the simplest case in which the interacting particles support just a single dipole mode[39-43]. Figs. 1c,d show the simulated field distribution when the array is illuminated by s- or p-polarized plane waves normally incident at the operating wavelength from the substrate side. We define the s-(p-) polarization as that with the magnetic (electric) field directed along the axis of the dimer. As seen from the figures, the energy is mainly concentrated into the $T_{+1}$ order. This is evidenced by the plane wave emerging from the array tilted at 82° with respect to its normal with minimal distortion. In the simulations, more than 88% of the transmitted energy is concentrated into the correct diffraction order for s-polarized light and more than 82% for p-polarized light.

The designed array was fabricated using conventional e-beam lithography technique on a quartz substrate and its diffraction efficiency was characterized as described in Methods. The simulated and measured spectral dependencies of diffraction efficiencies of this array, defined as the amount of power channeled into a particular order divided by the incident one, are provided in Fig. 2. The measured results, in good agreement with the simulated ones, show a diffraction efficiency into the $T_{+1}$ order of around 35% at 715nm for p-polarized light, representing around 75% of the transmitted power. For s-polarized light, the diffraction efficiency is slightly smaller, around 32%, representing 70% of the transmitted power being concentrated into the $T_{+1}$ order. While these values are still far from the ideal 100% efficiency, it should be mentioned that there is no fundamental limitation in the efficiencies achievable using the method described here. We foresee that further optimization of the nanoantenna scattering patterns could lead to designs beating the one reported here, mainly constrained by polarization insensitivity requirements, as has been recently theoretically reported using polarizability considerations[44].

*Demonstration of a near-unity numerical aperture metalens.*

On the basis of the concept presented above, a spherical lens is designed to have a numerical aperture of NA~0.99 at 715 nm wavelength. This corresponds to a maximum bending angle of 82°, realized with the system described above. The lens has a total diameter $d = 600$ μm and a focal distance $f \sim 42$ μm. The nanoantenna inclusions and unit cells on the different areas of the lens are designed to generate different bending angles corresponding to their radial positions within the lens, thus approximating the ideal parabolic phase profile to a piece-wise linear one, sampled in one-degree steps. All angles spanning from 82° to 55° are achieved with the asymmetric dimer design described above only changing the array period, with efficiencies that increase as the angle decreases. Angles ranging from 54° to 36° are achieved using asymmetric trimer nanoantennas. The diameters of the cylinders are $D_1 = 130$ nm, $D_2 = 150$ nm and $D_3 = 190$ nm and the gaps $g_{12} = g_{23} = 50$ nm (see Supplementary Fig. 1). Angles between 35° and 31° are achieved with quadrumers adding another disk with $D = 110$ nm (for 35° and 34°) and $D = 120$ nm (for 33° to 31°) and changing the inter-particle gaps to 60 nm (see Supplementary Fig. 2). In all cases the non-diffractive period is kept constant $P_{nd} = 260$ nm while the diffractive period $P_d$ is changed to generate the first diffraction order at the corresponding angle (the lens is discretized in Fresnel zones with an approximate angular step of 1°). Characterization of two examples, corresponding to arrays bending light at 40° and 35°, can be found in Supplementary Figs 1 and 2 in Supplementary Information, respectively. The values shown are representative of the whole range of angles, with diffraction efficiencies into the $T_{+1}$ order exceeding 60% at the operating wavelength, representing more than 85% of the total transmitted power. For angles below 30° the usual phase mapping approach is taken (with geometries and angular discretization specified in Supplementary Table 1). The central region of the lens, with a diameter $d_0 = 13.5$ μm is kept empty in this proof of concept demonstration.

The lens was fabricated using standard nanofabrication tools, namely, electron beam lithography followed by resist development and reactive ion etching of a 250 nm-thick amorphous Si film deposited on top of a quartz substrate using chemical vapor deposition (see Methods for details). Scanning electron microscope (SEM) images of the fabricated sample with different magnifications are shown in Figs. 3a-c. To characterize its focusing capability we use the experimental setup schematically depicted in Supplementary Fig. 3 and detailed in the Methods section. The results reveal the generation of a slightly elongated focal spot at a distance of approximately 42-47 μm from the lens, as shown in Figs. 3d and 3e. This implies a maximum experimental bending angle of around 82°, corresponding to a maximum NA of ~0.99. The elongated focus profile can be explained by ~10nm bandwidth of the filtered supercontinuum laser used for these measurements and associated dispersion effects (see Supplementary Table 2 and corresponding discussion below). The measurements also reveal a highly symmetric intensity distribution in the focal plane (xy-plane), shown in Fig. 3f, and a very low level of background. Figure 3g shows a cut across the x-axis passing through the focus, together with the fitting to the Airy pattern, corresponding to a diffraction limited system. The coefficient of determination of the fit is 0.93, which gives us a measure of the quality of the image formation. The measured FWHM along this cut is 385 nm, corresponding to an NA of around 0.935, which is at the limit of resolution of our measurement system. Note that, due to the limited numerical aperture of the objective ($NA_{obj}$) used to measure the focal spot (Nikon 100x, NA=0.95), for a diffraction limited system, the minimum FWHM measurable with this setup would ideally be $\lambda/2NA_{obj} \sim 376$ nm. This value, however, is not met in commercial lenses at all wavelengths, as was recently reported[30], in accordance with the slightly larger spot size measured in our experiment. Using the alternative definition based on the Airy radius obtained from the fitting the obtained NA is 0.944, in close agreements with the result obtained from the FWHM. We provide further proof of the limitation on the measured NA imposed by the commercial objective lens used comparing the back-focal plane images obtained through this objective lens from the focal spot created by the flat lens with that from a strongly scattering medium placed at its focus, and which generates a random distribution of wave-vectors. This comparison can be found in Supplementary Fig. 4 together with the back-focal plane images from focal spots generated by commercial objective lenses with NA=0.8 and 0.9.

By integrating the intensity distribution in the focal plane (we restrict the integration of the intensity to the area shown in Fig. 3f) we found that only around 10% of the incident power is focused by the lens. We attribute this small value to the non-optimized packing density of the bending arrays within the circular lens and to observed deviations in the nanoparticles sizes from the designed ones, causing the drop in efficiency as compared to simulations. Both of these factors, however, may be circumvented with appropriate design and fabrication corrections and we feel, nevertheless, that the values obtained are reasonable for a proof-of-concept system. A comparison of the focusing characteristics of fabricated sample and a commercial objective lens (Zeiss x100, NA 0.9) can be found in Supplementary Fig. 5, showing a clearly smaller FWHM of the focal spot in the case of the presented flat lens and comparable focus quality characteristics.

In order to prove that the NA of the fabricated flat lens indeed reaches its designed value of 0.99 we perform an additional experiment in which the lens is used to collect light from a single sub-diffractive scatterer located at its focus. In the experiment, the scatterer is a small dust particle deposited on top of a triangular prism. The particle is excited by the evanescent fields generated in a total internal reflection configuration, as shown in Fig. 4a. In this situation, and upon unpolarized excitation, the dust particle generates scattering into all directions and, thus, the full aperture of the lens can be proved while avoiding any other source of illumination into the lens. When the particle is placed at the focus of the metalens and illuminated at the operating wavelength (715 nm), the scattered light collected by the metalens emerges as a collimated beam propagating parallel to its optical axis, as shown schematically in Fig. 4a. This is subsequently imaged using a low NA objective (Nikon 10x, NA=0.3) to collect, in good approximation, only those rays that are parallel to the optical axis. The collected normalized intensity distribution is shown in Fig. 4b. The high intensity area in the image coincides with the physical size of the lens, therefore demonstrating that the full area of the lens is functional. Similar images were collected at different distances from the lens showing similar results and therefore corroborating that the emergent beam is fully parallel to the optical axis. This, in turn, demonstrates that the lens aperture coincides with its physical size and that the corresponding NA meets indeed the design value of 0.99. As a comparison, we use the metalens to collect light from the focal spot generated by a commercial objective lens with NA 0.95. In this situation only a part of the lens corresponding to collection angles < 72º is functional/illuminated, the size of this illuminated area being much smaller than the full size of the lens, in very good agreement with theoretical estimations. These results, presented in comparison with those from the small scatter, can be found in Supplementary Fig.6. Additionally, an alternative experimental setup and set of measurements, in which diffraction from small holes on a metal screen is collected instead, can be found in Supplementary Fig. 7. This experiment serves to further corroborate the results obtained using the small scatter.

*Fluorescence imaging of solid-state quantum emitters using the metalens.*

Finally, to check the imaging capabilities of the designed metalens, we characterize the photoluminescence (PL) from nitrogen-vacancy (NV) centers in nanodiamonds (NDs) in the scanning confocal configuration shown in Fig.5a (see Methods for details). This sample was chosen since NV centers possess their zero-phonon lines at 575 nm and 637 nm with a broad phonon sideband spanning the range from 550 nm to 800 nm, which includes the operating wavelength of the metalens (see Supplementary Fig.8). In addition to collecting emitted light, the metalens acts as a band-pass filter, selecting different wavelengths as a function of the distance from the ND. When the ND is placed exactly at the designed focus, the PL passing through the metalens presents a peak at 715 nm, precisely coinciding with the designed operating wavelength (see Supplementary Figs.8 and 9 and Supplementary Table 2). In the experiment, the confocal maps are performed, for comparison, both in reflection using the same commercial objective that is used for illumination (Mitutoyo, NA=0.70) and in transmission using the metalens. Prior to the confocal mapping, SEM characterization of the deposited sample was done to find isolated NDs (or small ND clusters) with size less than 100 nm. The inset in Fig. 5b shows an SEM image of the measured ND, in this case forming a cluster with a size below 90 nm. This area was traced back by the confocal microscope and characterized using both the commercial objective and the metalens. The confocal scans at 715nm of an area of 20 μm x 20 μm around this ND performed through the commercial objective (NA=0.7) and the metalens are shown in Figs. 5b and 5c respectively. Both confocal maps are very similar, indicating that, even if far from optimal, this metalens can be used to collect the weak optical signal from the small quantum emitters.

Figures 5d and 5e show fine confocal scans of the 90 nm ND circled in Figs. 5b and 5c and imaged by SEM (inset in Fig.5b). Each scan was repeated 4 times and the results were averaged to obtain the images. The results show, clearly, a smaller spot in the case of the scan performed using the metalens as compared to that retrieved using the commercial objective. The reconstructed spot shows a fairly good signal to noise ratio, confirming the imaging capability of the metalens in the proposed confocal configuration and representing, to the best of our knowledge, the first example in which a metalens is used for fluorescence imaging of a solid-state emitter.

**Discussion**

The novel concepts presented here, in which nanoantennas with tailored scattering patterns are used to engineer the redistribution of the diffracted energy in periodic arrays, may find important applications in the field of optical gratings, where a precise control of energy distribution is crucial. Moreover, the associated flat lens design, with a near-unity numerical aperture, may be useful in a wide range of applications requiring high resolution focusing, such as photolithography. On the other hand, by reciprocity arguments, the presented lens may find important applications in those situations requiring large angular collection, such as quantum optics experiments involving isotropic emission of a small number of photons. There, this apparently small increase in the NA of the lens translates in a truly significant increase of the collection solid angle, which will boost the efficiency of the mentioned processes (see one-to-one comparison of the collection area of the flat lens and a commercial objective with NA=0.95, highest available at the market, in Supplementary Figure 6). This would require further optimization of the lens fabrication and design to reach efficiencies comparable to commercial microscope objectives.

Overall, we believe that this work represents a clear example of how resonant nanoantennas may help to overcome the limitations of phase mapping approaches to diffractive optics, and how metasurface concepts may lead to truly sub-wavelength-thick flat optical devices that break the limitations of their traditional counterparts.

**Methods**
*Sample fabrication*
    Thin films of amorphous silicon (250 nm thickness) were deposited on fused silica substrates via chemical vapor deposition (Oxford PECVD). The samples were then patterned

using single-step electron beam lithography (Elionix ELS-7000) using Hydrogen silsesquioxane (HSQ, Dow Corning XR-1541-006) resist and a charge-dissipation layer (Espacer 300AX01), and subsequently etched via reactive-ion-etching in inductively coupled plasma system (Oxford Plasmalab 100) using chlorine gas.

*Optical measurements*

Two types of optical measurements were carried out to characterize the diffraction efficiencies in transmission of uniform arrays of Si disks. Spectrally-resolved back focal plane imaging measurements, used to characterize the diffraction efficiencies of low angle bending samples, were done using an inverted microscopy setup (Nikon Ti-U). In these measurements, white light from a halogen lamp was used to normally illuminate the sample from the substrate side. Polarization was selected using a linear polarizer (Nikon). A spectrometer (Andor SR-303i) with an EMCCD (Andor Newton) detector was placed in the back focal plane of the collection objective (Nikon, x100, NA 0.9) to measure the transmitted spectrum as well as the k-vector distribution, as described in (26). For measuring large-angle-bending samples, for which the diffraction angles were larger than the maximum acceptance angle of the collection microscope objective in our back focal plane measurement setup, a home built free-space microscopy setup was used. A supercontinuum source (SuperK Power, NKT Photonics) and band-pass filter (SuperK Varia, NKT Photonics) were used to illuminate the sample. The beam was transmitted through a broadband polarizing beam-splitter cube (Thorlabs, PBS252). The linearly polarized light passed then through a quarter wave plate (Thorlabs, WPQ10M-808) to obtain circularly polarized light, which was sent to a rotating linear polarizer (Thorlabs, LPNIRE100-B) to obtain light with the desired linear polarization. Light was then focused onto the sample using a biconvex lens (Thorlabs) with a 75 mm focusing distance. The diffraction orders were then measured using a pixel photodetector attached to a digital handheld laser power/energy meter console (Thorlabs, PM100D) mounted in a rotating arm, as explained in detail elsewhere[45].

To characterize the focusing capabilities of the fabricated flat lens we used the experimental setup schematically depicted in Supplementary Fig. 3. Light generated with the same supercontinuum source was transmitted through the variable band pass filter for wavelength selection. The beam was then expanded using a beam expander (Melles Griot, x10) and directed to the sample using several mirrors. Light focused with the metalens sample was collected using an objective lens (Nikon x100, NA 0.95) and imaged to a CCD detector (Qicam Fast). The sample was mounted on a piezo stage that allowed measuring the intensity maps at the different planes from the sample.

*Confocal measurements*

NDs with a size of 25 nm (Diamond Technologies, Inc.) were drop cast onto a quartz substrate, mounted on a 3D piezoelectric stage (Piezosystem Jena, TRITOR 100 SG) and scanned in the confocal configuration shown in Fig.5a with a 20 nm to 500 nm step depending on the scan range. Excitation light from a continuous-wave 532nm laser (Oxxius) was focused onto the sample through an air objective (Mitutoyo, NA=0.70) with an excitation power of 570 $\mu$W. The emission was filtered by a dichroic mirror (Semrock FF555-Di03), a notch filter (Semrock NF03-532E), and a bandpass filter (Semrock FF01-697/75-25-D). The emission was subsequently collected using a multimode fiber (62.5 $\mu$m core) and directed to either a grating spectrometer (Princeton Instruments IsoPlane 160, 0.07nm resolution) with the narrowband filter removed for PL measurements or an avalanche photodetector (APDs, Perkin Elmer SPCM-AQRH-15) to perform the scanning confocal map, as depicted in Fig.5a.

For the transmission measurement, both excitation and emission should be well-focused onto the sample. Thus a confocal scan was first performed with the commercial objective to ensure a proper focusing of the excitation light. Then we measured the transmission through the metalens. For the accurate measurement of the focal distance, another z-axis piezoelectric stage was placed on the metalens, which enabled measuring the focal distance with down to 10 nm accuracy.

*Numerical simulations*

Full wave numerical simulations of diffraction efficiencies of regular arrays of nanoantennas were performed using a Finite Element Method (FEM) based commercial software (Comsol Multiphysics V5.2). The simulation domain consisted of a single unit cell. The unit cell was divided into two subdomains, top and bottom, representing, respectively, the surrounding medium of the antennas (air) and the substrate (glass, modeled as a medium with constant refractive index $n_s$=1.46). The nanoantennas were modeled as cylinders with the appropriate dimensions with material parameters taken from experimental measurements[46]. Bloch boundary conditions were applied in the lateral sides (x and y directions) of the system to simulate an infinite rectangular lattice. Several port boundary conditions were applied in the vertical directions (z direction) to collect the diffracted orders both in transmission and reflection. One of the ports (on the substrate side) was used to excite the system with a plane wave with appropriate polarization. The injection angle was set to zero to simulate a normally incident wave.

All FEM simulations presented were checked by simulating the same system using a Finite Difference Time Domain (FDTD) based commercial software (Lumerical FDTD Solutions). The same results were obtained within an acceptable numerical error.

## Acknowledgments

We acknowledge Marcus Cheong and Indra Gunawan for their help in setting up and operating the confocal setup. The authors acknowledge support from DSI core funds, A*STAR SERC Pharos program, Grant No. 152 73 00025 (Singapore) and National Research Foundation grant No. NRF-CRP14-2014-04 (Singapore).

## Author contributions

R. P.-D. and A. I. K. conceived the idea. E. K. designed the nanoantennas and R. P.-D. designed the flat lens. Both E. K. and R. P.-D. performed the numerical simulations. Y. F. Y. fabricated the samples and performed the optical measurements. S. C., V. L. and L. A. K. built the confocal setup and performed fluorescence measurements. Y. H. F. and X. L. helped with the optical measurements. V. V. performed the SEM imaging and Cr holes fabrication. R. M. B. performed NSOM measurements to corroborate the measured focal spot. R. P.-D. wrote the first draft of the manuscript. A. I. K. coordinated the work. All authors discussed the results and reviewed the manuscript.

## Competing financial interests

The authors declare no competing financial interests.

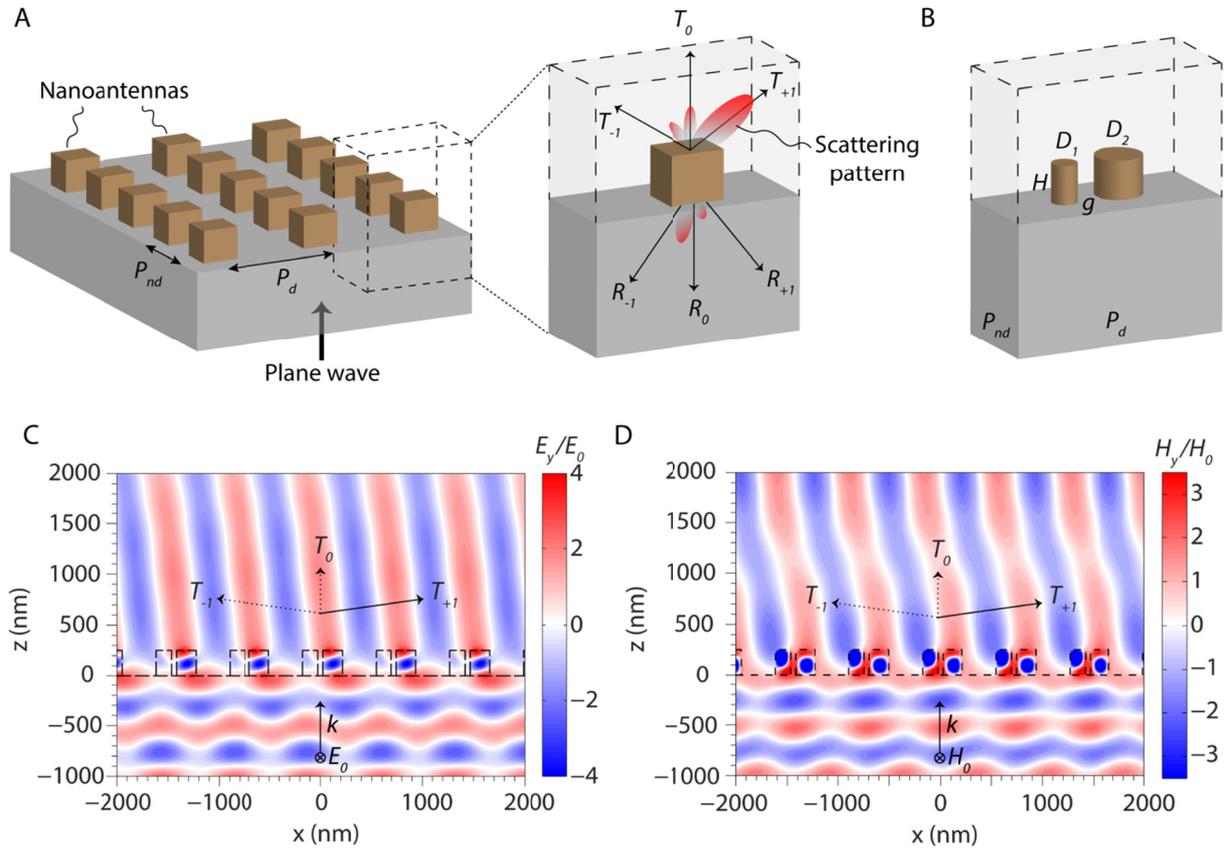

**Fig. 1. Operating mechanism of the large angle light bending arrays used in the near-unity NA lens design.** (**A**) Schematic description of the nanoantenna arrays with controlled energy distribution among the supported diffraction orders. The number and angle of these orders is determined by the diffractive period of the array, $P_d$. The energy redistribution is determined by the scattering pattern of each nanoantenna in the array. (**B**) Schematic description of the array of asymmetric dimers producing energy concentration into the T+1 diffraction order leading to light bending at 82° for plane waves normally incident from the substrate side at a wavelength of 715 nm. (**C**) - (**D**) Simulated out-of-plane component of the electric (C) and magnetic (D) field when the array is illuminated by a normally incident plane wave with the magnetic (C) or electric (D) field along the axis of the dimer.

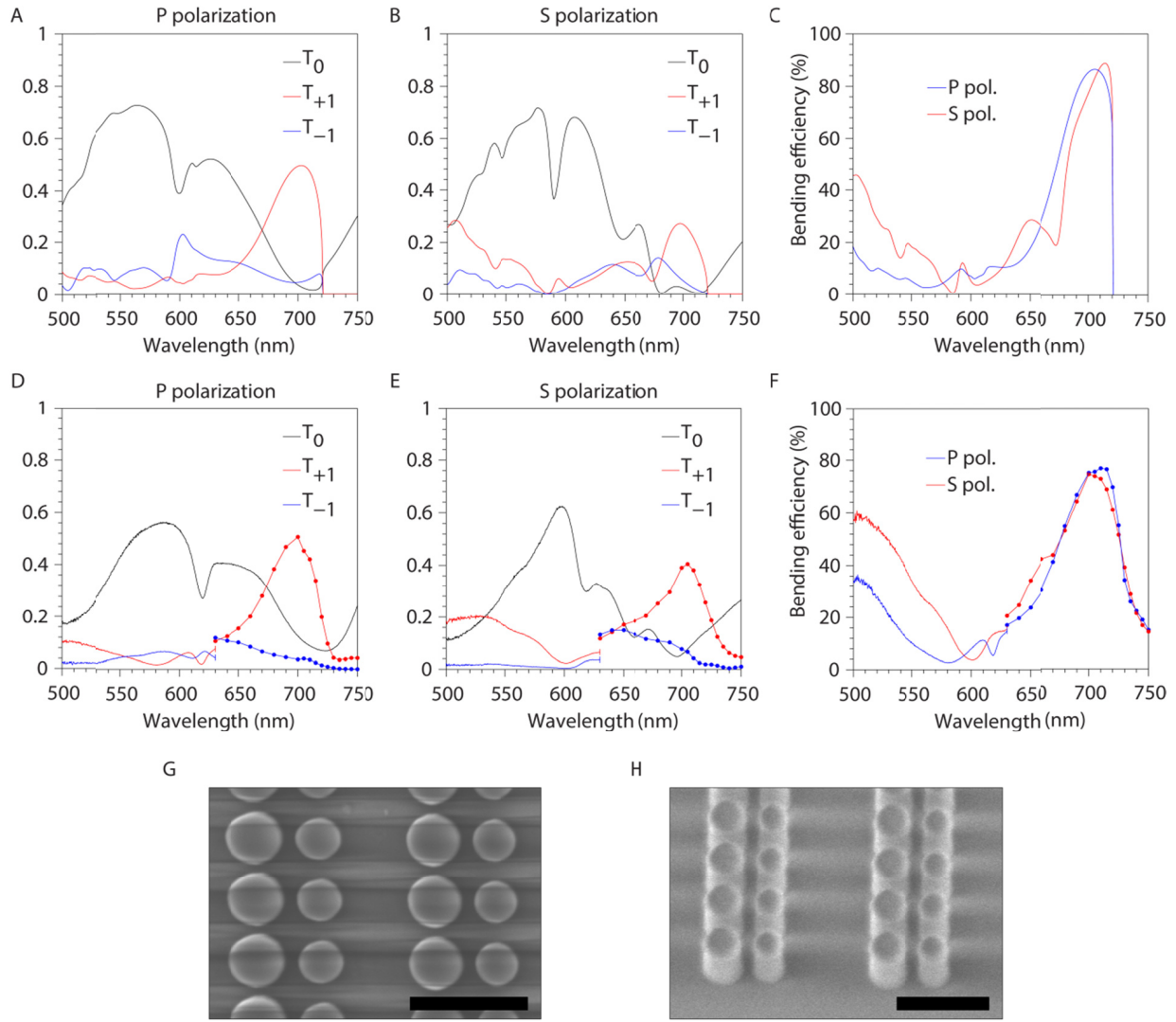

**Fig. 2. Numerical and experimental characterization of silicon dimer arrays bending light at 82 degrees. (A)** to **(B)** Simulated diffraction efficiencies under normally incident light illumination with (A) p- or (B) s-polarization as defined in the main text. **(C)** Simulated bending efficiency of the array, representing the amount of power into the desired order relative to the total transmitted one. **(D)** to **(E)** Measured diffraction efficiencies under normally incident light illumination with (D) p- or (E) s-polarization. The solid lines correspond to back-focal plane measurements while those with symbols correspond to measurements carried out using a free-space microscopy setup, as explained in Methods. **(F)** Measured bending efficiency of the array. **(G)** to **(H)** SEM images of the fabricated sample under top (G) and tilted (H) views. The scale bars correspond to 500 nm.

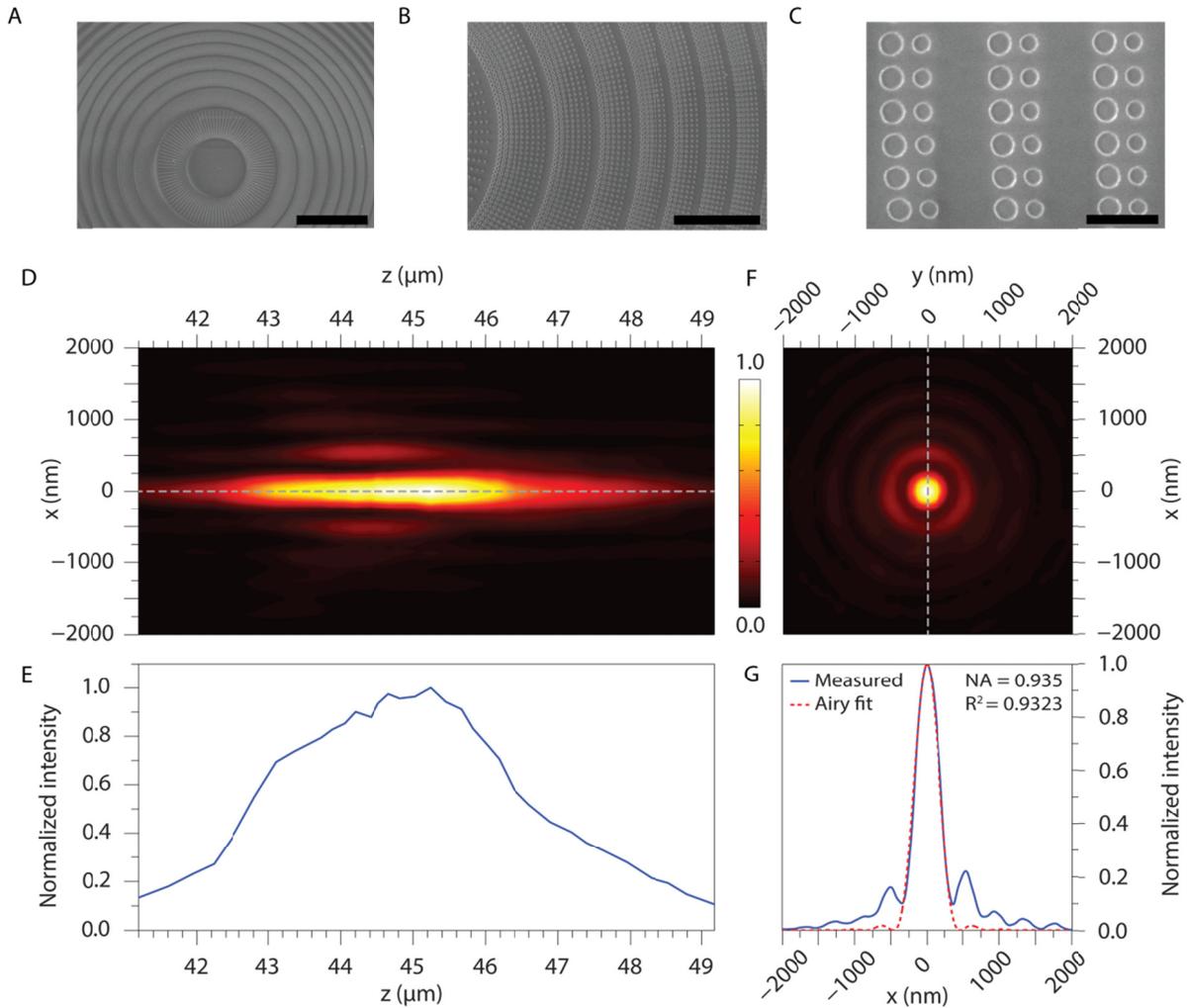

**Fig. 3. Experimental characterization of the high-NA flat lens.** (**A** to **C**) SEM images of the fabricated sample showing (A) a low magnification image of the central part, (B) a low magnification view of the low angle bending part of the lens and (C) a large magnification view of the large angle bending part of the lens. The scale bars represent 10 μm, 5 μm, and 500 nm, respectively. (**D**) Normalized measured intensity map along the x-axis as a function of the distance to the lens, z, around the lens focus. (**E**) Normalized measured intensity profile along the axis of the lens as a function of the distance, z, around the lens focus. (**F**) Normalized measured intensity map in the focal plane (z = 45 μm). (**G**) Intensity profile along the x-axis, passing through the focus, together with the fitting to the Airy profile.

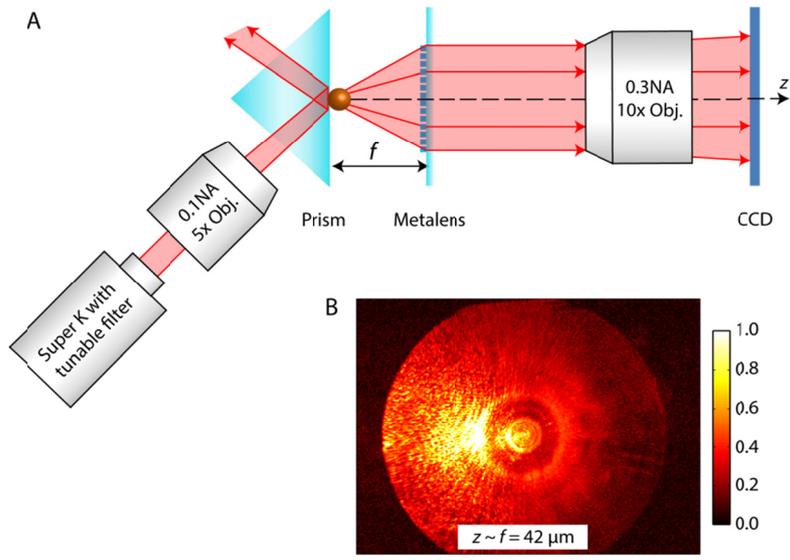

**Figure 4. Demonstration of the numerical aperture of the metalens by collecting light from a single, sub-diffractive particle located at its focus.** (**A**) Schematic of the experimental setup. (**B**) Normalized intensity of the collimated beam emerging from the metalens. The physical size of the lens coincides with the high intensity area in the image.

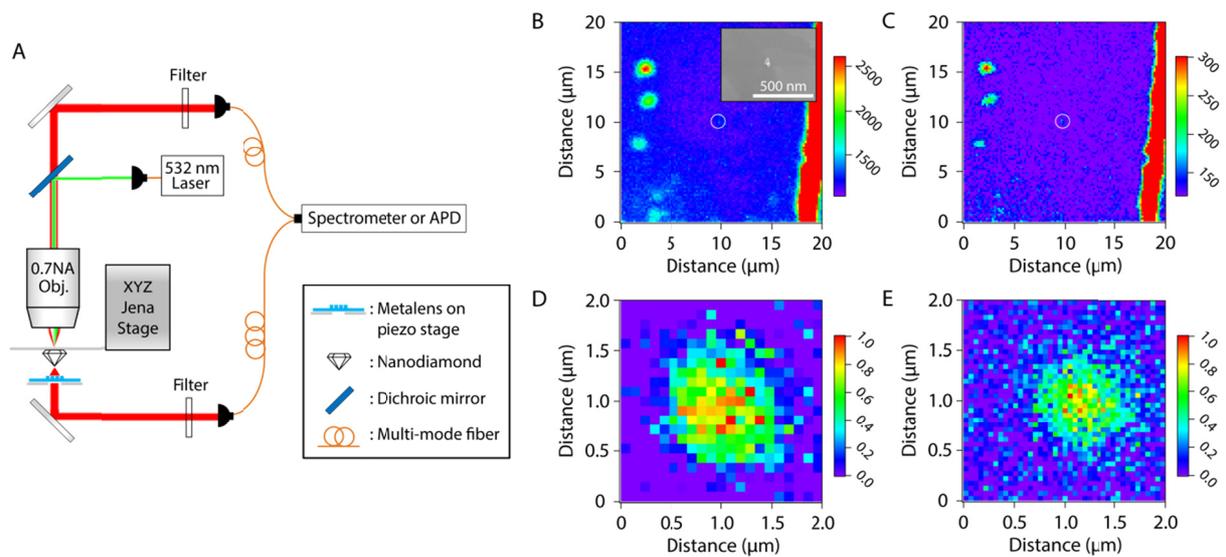

**Figure 5. Confocal map of photoluminescence (PL) from nanodiamonds (NDs) retrieved using the metalens.** (**A**) Schematic of the experimental setup. (**B**)-(**C**) Confocal maps of 20 μm × 20 μm using (B) a commercial objective with NA=0.7 and (C) the metalens. The step size is 0.2 μm and the color scale bar denotes count rates per 50 ms. The inset in panel (B) shows an SEM image of the ND circled in (B)-(C) and further characterized by the optical measurement. (**D**)-(**E**) Normalized zoom-in confocal maps of the circled ND using (D) the commercial objective lens and (E) the metalens. The step sizes are 100 nm for the commercial objective and 50 nm for the metalens, respectively.

# Supplementary Information - A Metalens with Near-Unity Numerical Aperture

Ramón Paniagua-Domínguez[1†]*, Ye Feng Yu[1†], Egor Khaidarov[1, 2†], Sumin Choi[1], Victor Leong[1], Reuben M. Bakker[1], Xinan Liang[1], Yuan Hsing Fu[1], Vytautas Valuckas[1], Leonid A. Krivitsky[1] and Arseniy I. Kuznetsov[1#]

[1]Data Storage Institute (Agency for Science, Technology and Research, A*STAR), 2 Fusionopolis Way, #08-01, Innovis 138634, Singapore.

[2]LUMINOUS! Center of Excellence for Semiconductor Lighting and Displays, The Photonics Institute, School of Electrical and Electronic Engineering, Nanyang Technological University, 50 Nanyang Avenue, 639798, Singapore

*Correspondence to:  ramon-paniag@dsi.a-star.edu.sg.

#Correspondence to:  arseniy_k@dsi.a-star.edu.sg.

†These authors contributed equally to this work.

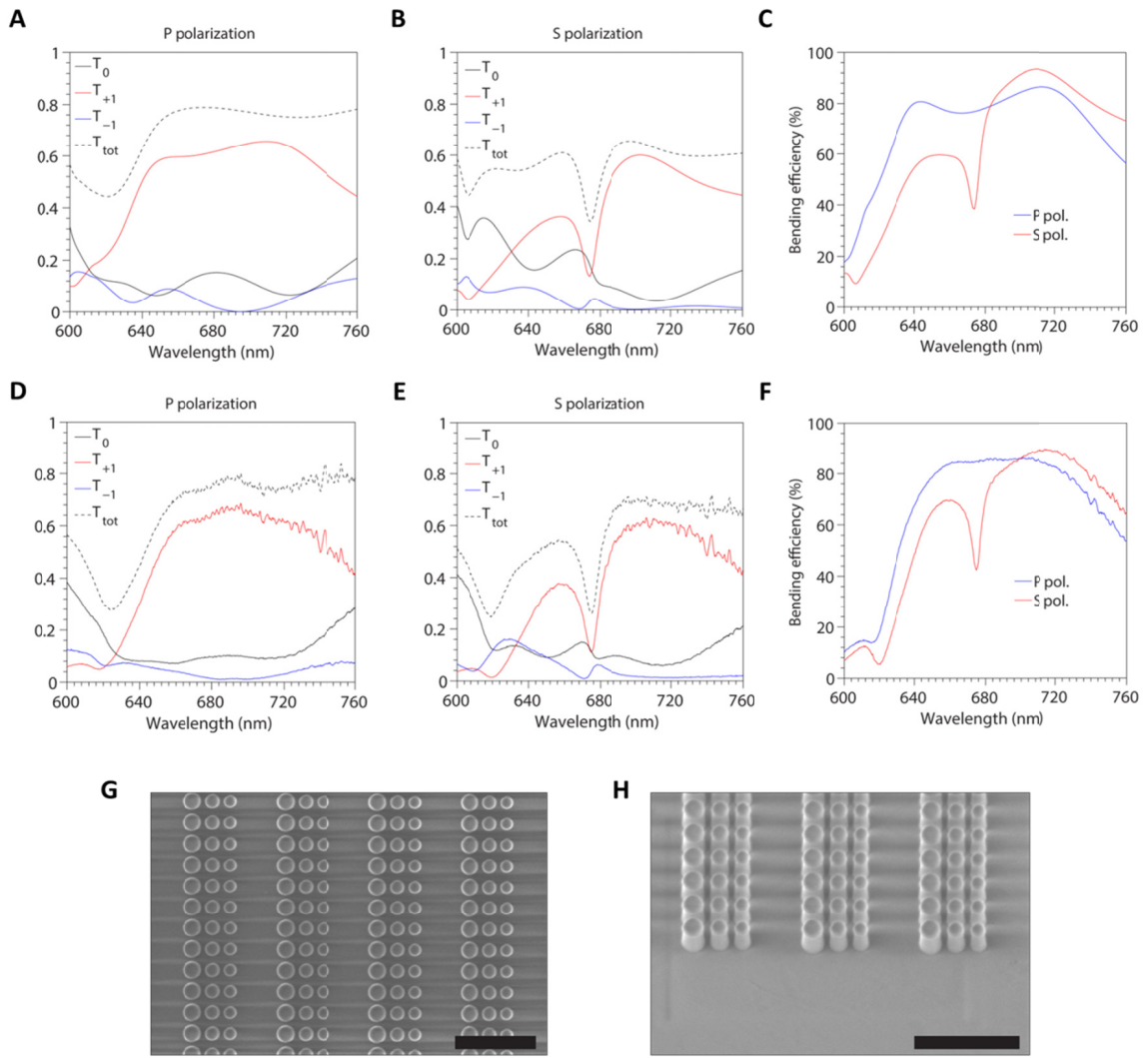

**Supplementary Fig. 1. Numerical and experimental characterization of silicon trimer arrays bending light at 40 degrees.** (A to C) Simulated diffraction efficiencies under normally incident light illumination with (A) p- or (B) s-polarization as defined in the main text and (C) simulated bending efficiency of the array, representing the amount of power into the desired order relative to the total transmitted one. (D to F) Corresponding back focal plane measurements of the diffraction efficiencies under (D) p- or (E) s-polarization and (F) measured bending efficiency of the fabricated sample. (G to H) SEM images of the measured samples from top (G) and tilted views (H). The scale bars correspond to 1μm.

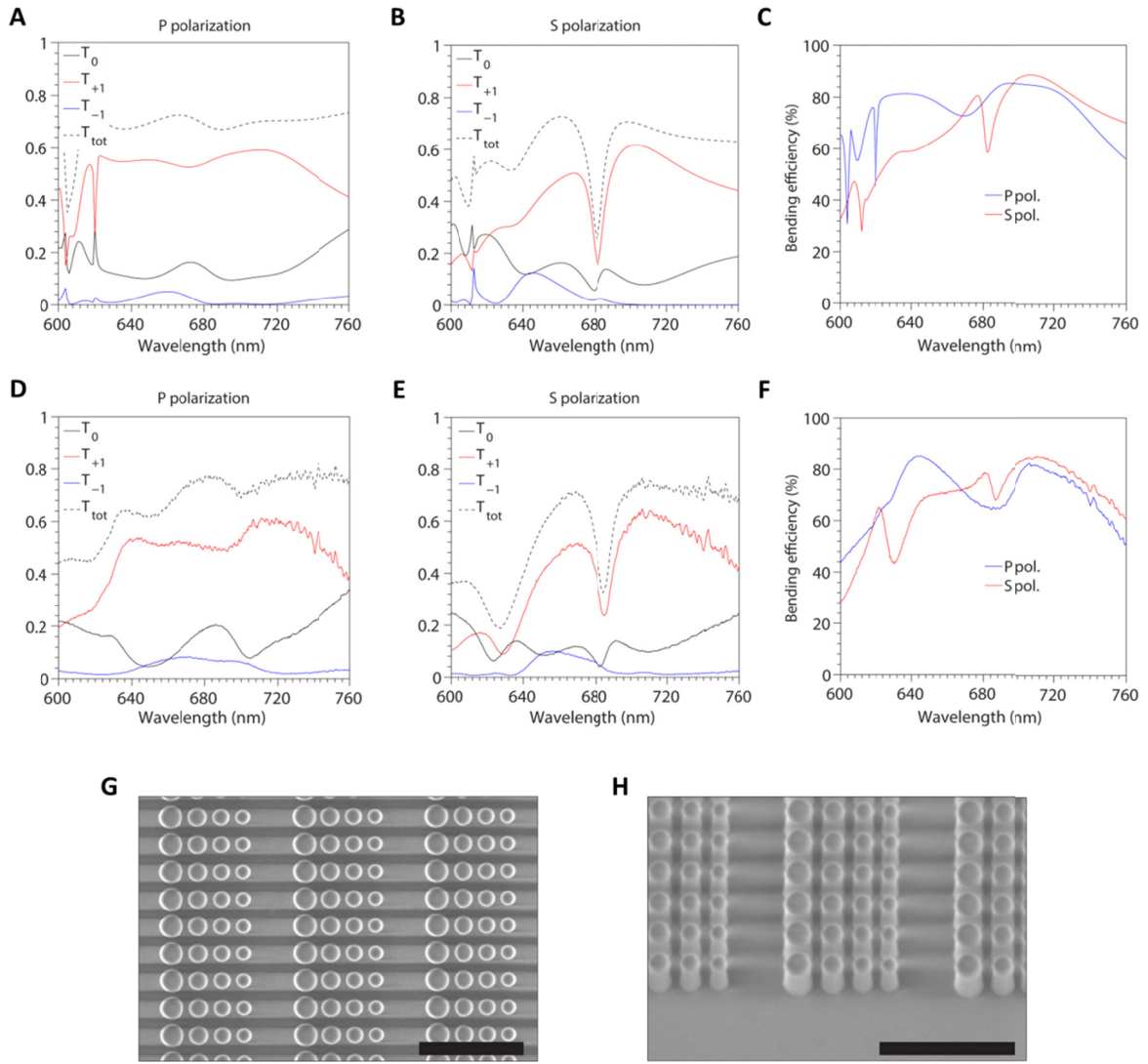

**Supplementary Fig. 2. Numerical and experimental characterization of silicon quadrumer arrays bending light at 35 degrees.** (A to C) Simulated diffraction efficiencies under normally incident light illumination with (A) p- or (B) s-polarization as defined in the main text and (C) simulated bending efficiency of the array, representing the amount of power into the desired order relative to the total transmitted one. (D to F) Corresponding back focal plane measurements of the diffraction efficiencies under (D) p- or (E) s-polarization and (F) measured bending efficiency of the fabricated sample. (G to H) SEM images of the measured samples from top (G) and tilted views (H). The scale bars correspond to 1μm.

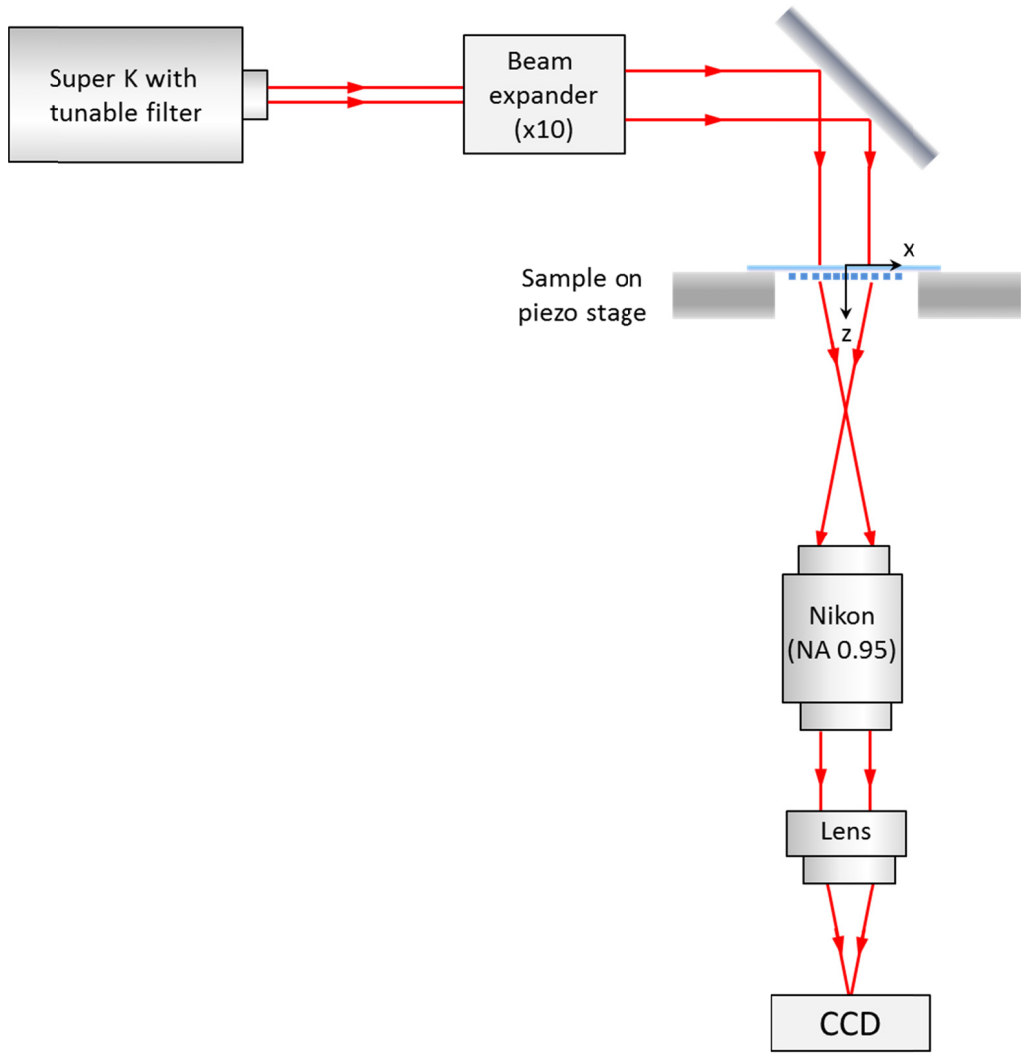

**Supplementary Fig. 3. Experimental setup used for the optical characterization of the fabricated flat lens.**

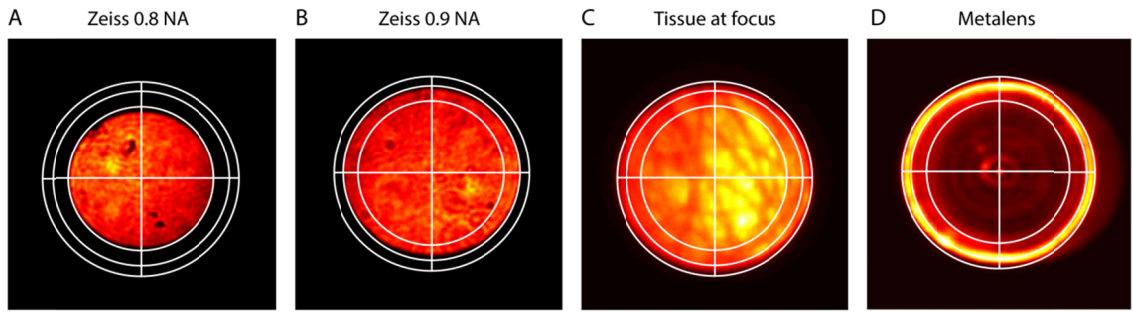

**Supplementary Fig. 4. Comparison of the back-focal plane images collected with a commercial objective lens with a NA=0.95 (Nikon x100) from different sources placed at its focus.** (a) Back-focal image from a focal spot generated with a commercial objective lens with NA=0.8 (Zeiss 50x). (b) Back-focal image from a focal spot generated with a commercial objective lens with NA=0.9 (Zeiss 100x). (c) Back-focal image from a strongly scattering medium (a tissue) placed at the focus of the objective. (d) Back-focal image from a focal spot generated with the proposed metalens. In both cases (c) and (d) the generated k-vectors reach the limit of collection of the 0.95 NA commercial lens. The bright ring in (d) corresponds to the signal coming from a large lens area responsible for high-angle bending and squeezed into a narrow ring in the back focal plane image.

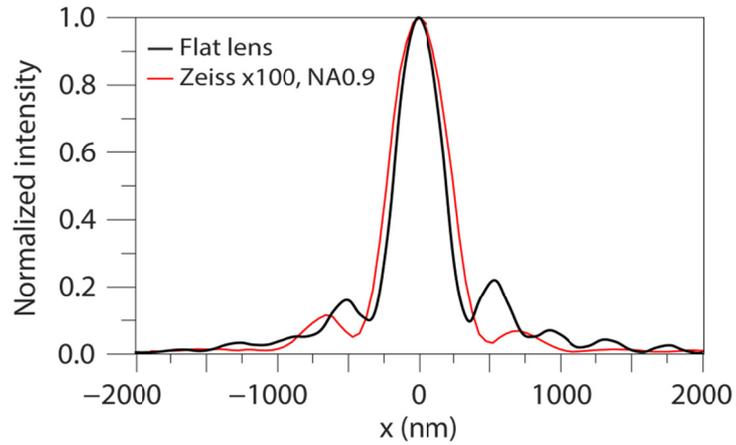

**Supplementary Fig. 5. Comparison of the focus of the fabricated flat lens with that of a commercial objective lens (Zeiss 100x, NA 0.9).** The plot shows the normalized intensity profile along the x-axis passing through the focal spot (the measurements were performed with a commercial objective lens Nikon x100 with NA=0.95).

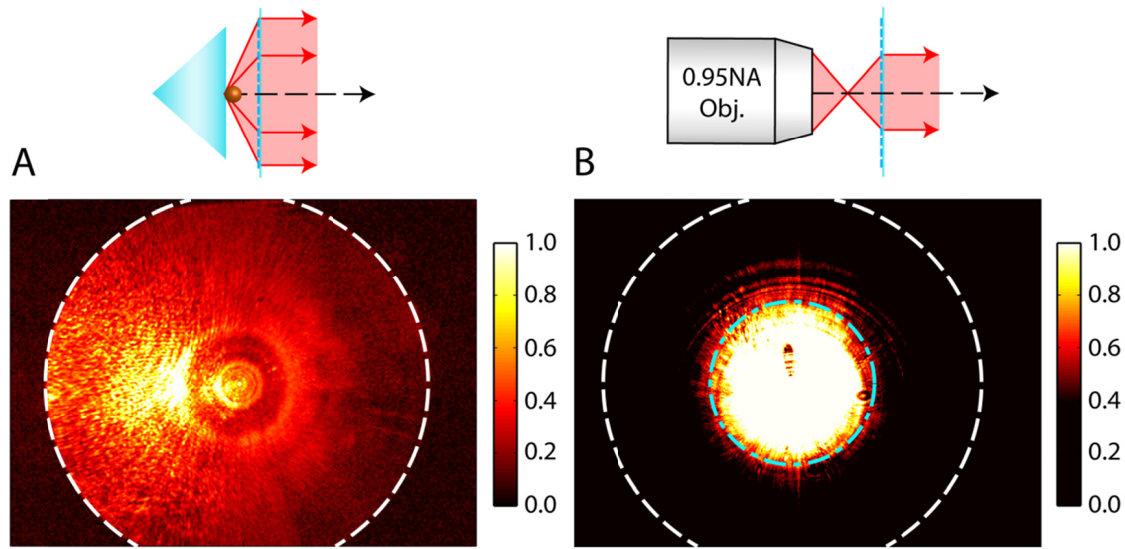

**Supplementary Fig. 6. Comparison of the functional area (numerical aperture) of the metalens when it is used to collect light from a small scatterer located at its focus and from the focal spot generated using a commercial objective lens with NA=0.95.** (**A**) Normalized intensity of the beam emerging from the metalens when collecting light from a small sub-diffractive scatterer located at its focus. (**B**) Normalized intensity of the beam emerging from the metalens when collecting light from the focal spot generated by a commercial objective lens with NA=0.95 (Nikon 100x).The white dashed lines in both plots indicate the edge of the metalens (corresponding to NA=0.99). The cyan, dot-dashed line in panel (B) represents the edge of the functional area corresponding to NA=0.95 calculated using simple geometric optics.

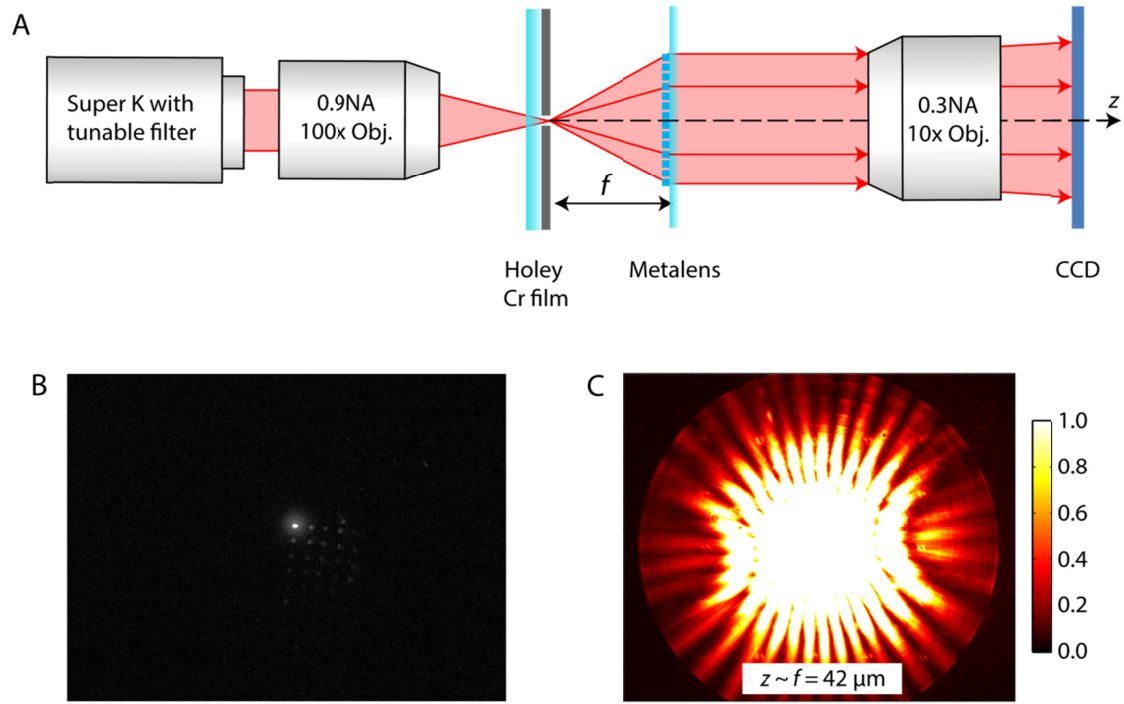

**Supplementary Fig. 7. Demonstration of the numerical aperture of the metalens by collecting light diffracted from an array of holes located at its focus.** (**A**) Schematic of the experimental setup. (**B**) Optical image of the array of holes. The light is tightly focused on one of the holes in the array. (**C**) Normalized intensity of the beam emerging from the metalens. The intensity is a convolution of the light scattered by a single hole and the effect of the array, as evidenced by the diffraction pattern observed. From the image, the different Fresnel zones of the lens can be observed as concentric rings at the edge of the lens.

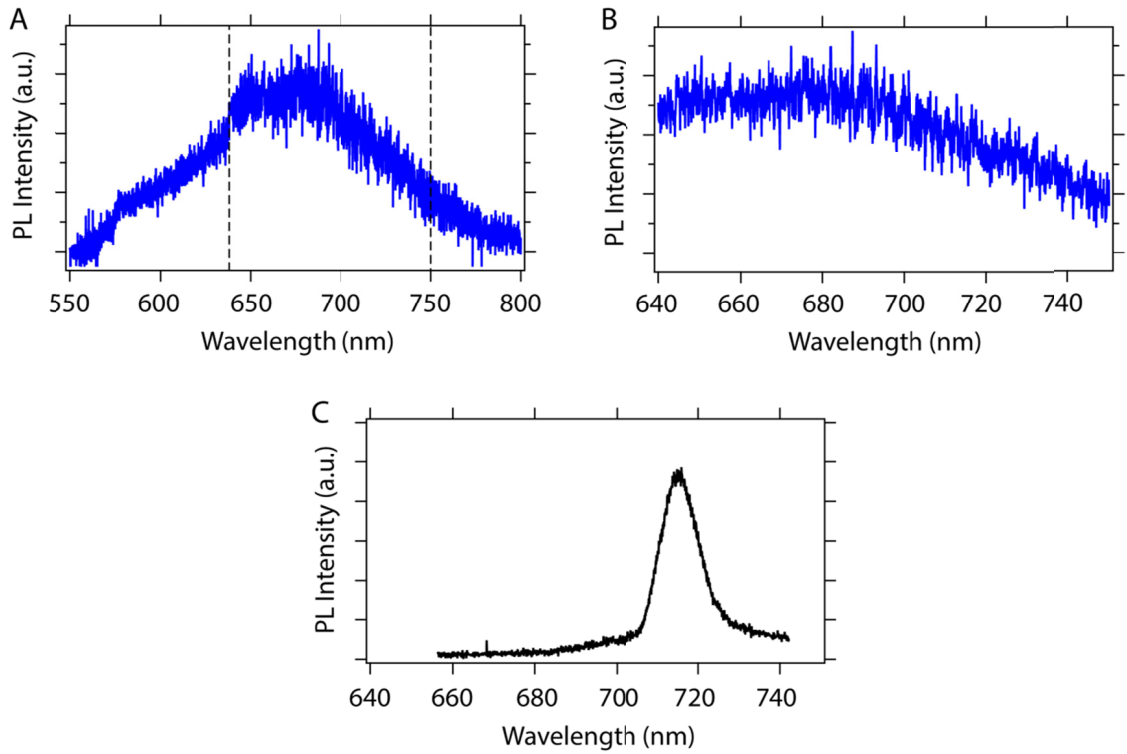

**Supplementary Fig. 8. Photoluminescence (PL) spectra from the measured nitrogen-vacancy (NV) center in a nanodiamond.** (**A**) Full spectral range of PL from the NV center in a ND collected using a commercial objective lens (Mitutoyo, NA=0.70) . (**B**) Same as (A) but in the reduced spectral range around the operating wavelength of the metalens. (**C**) PL spectrum as collected with the metalens when the nanodiamond is placed at its focus.

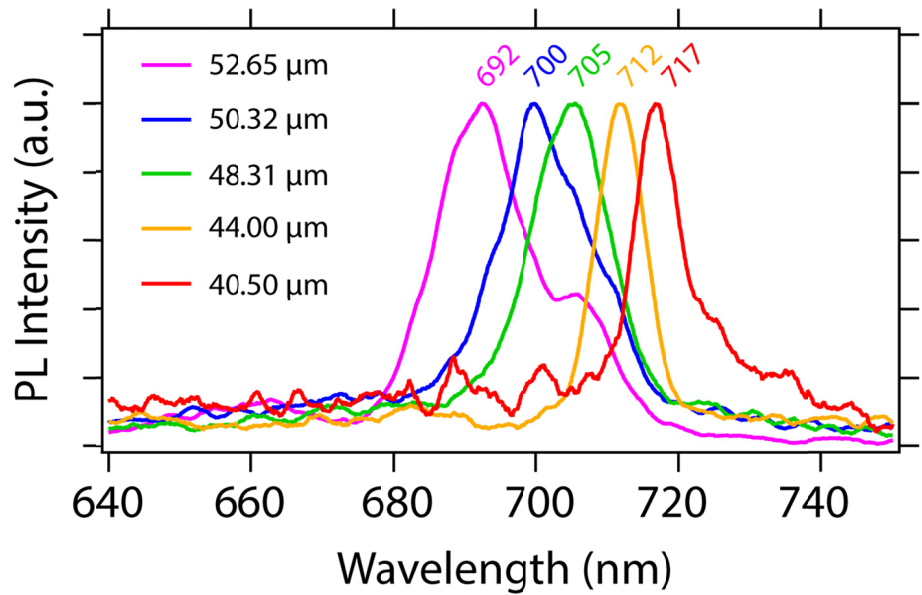

**Supplementary Fig. 9. PL spectra measured through the metalens from an NV center in a nanodiamond as a function of its distance to the metalens.**

| Θ (°) | P_tot(nm) | P_d (μm) | D_1(nm) | D_2(nm) | D_3(nm) | D_4(nm) | D_5(nm) | D_6(nm) | D_7(nm) | D_8(nm) | D_9(nm) | D_10(nm) | D_11(nm) | D_12(nm) | D_13(nm) | D_14(nm) | D_15(nm) | D_16(nm) |
|---|---|---|---|---|---|---|---|---|---|---|---|---|---|---|---|---|---|---|
| 9 | 283 | 4.56 | 232 | 219 | 207 | 197 | 189 | 183 | 177 | 171 | 165 | 158 | 153 | 146 | 139 | 128 | 108 | 50 |
| 14 | 280 | 2.95 | 224 | 205 | 191 | 180 | 171 | 161 | 151 | 141 | 122 | 50 | | | | | | |
| 17 | 277 | 2.44 | 227 | 204 | 186 | 170 | 156 | 142 | 119 | 50 | | | | | | | | |
| 23 | 261 | 1.827 | 183 | 210 | 197 | 179 | 163 | 146 | 123 | 50 | | | | | | | | |
| 25 | 282 | 1.689 | 219 | 190 | 170 | 152 | 128 | 50 | | | | | | | | | | |
| 27 | 262 | 1.572 | 212 | 190 | 170 | 152 | 128 | 50 | | | | | | | | | | |
| 28 | 253 | 1.518 | 202 | 190 | 170 | 152 | 128 | 50 | | | | | | | | | | |
| 30 | 285 | 1.425 | 198 | 178 | 171 | 140 | 0 | | | | | | | | | | | |

**Supplementary Table 1. Table with the different designs used in the central part of the flat lens.**

| Distance (um) | PL peak position (nm) |
|---|---|
| 38 | 719.3 |
| 39 | 718.8 |
| 40 | 717.6 |
| 41 | 716.4 |
| 42 | 715.0 |
| 43 | 713.6 |
| 44 | 712.1 |
| 45 | 710.3 |
| 46 | 709.2 |
| 47 | 707.1 |

**Supplementary Table 2. Peak positions of the PL spectra as a function of the distance between the nanodiamond and the metalens around the designed focus of the lens.**